\documentclass[showpacs, twocolumn, amsmath, amsfonts, prb]{revtex4}
\usepackage{graphicx}
\usepackage{graphics}
\usepackage{bm}
\usepackage{dcolumn}
\usepackage{amsmath}
\usepackage{amssymb}
\begin{document}

\title{Inelastic quantum tunneling through disordered potential barrier}
\author{Alex Levchenko}
\affiliation{Department of Physics, University of Minnesota,
Minneapolis, MN 55455, USA}
\begin{abstract}
The effect of inelastic scattering on quantum tunneling through a
rectangular potential barrier, of length $L$, containing randomly
distributed impurities, is considered. It is shown that, despite
the fact that the inelastic transition probability
$\mathcal{T}_{\mathrm{inelastic}}$ is exponentially small, it may
be greater than that of the elastic in the out of resonance
conditions for some energy interval $\triangle E$ (the energy of
the incident particle is far from the spectrum of the random
Hamiltonian). The asymptotic of the tunnelling transition
probability for various impurity configurations are found and the
energy interval $\triangle E$ where
$\mathcal{T}_{\mathrm{elastic}}\sim\exp(-L/L^{*})\lesssim\mathcal{T}_{\mathrm{inelastic}}$
is determined.
\end{abstract}
 \maketitle
\section{Introduction}
Due to advances in manufacturing techniques of semiconductor
devices, that make it possible to create small tunnel junctions,
and the need to understand fundamental conduction processes in
order to improve these devices, quantum-mechanical tunneling of
electrons through single, double or multi-barriers disordered
resonant tunneling structure, is in the focus of physical interest
Ref.~\cite{LGP}-\cite{FrelYurk} and \cite{Misc1}-\cite{Misc2}. The
most important peculiarity of these systems is that some impurity
configurations may increase the tunneling probability
dramatically. At the same time, the change of any single localized
state will change the magnitude of the resonant transmission
coefficient or even destroy the conditions needed for the resonant
tunneling. There is the general belief that resonant tunneling via
a single localized state (or a single quantum well) can dominate
the whole tunneling process. Assuming this, the resonant tunneling
through the whole system can approximately be treated using the
resonant tunneling via a single localized state. Then, the
Breit-Wigner formula is postulated to express the transmission
coefficient as
\begin{equation}
\mathcal{T}(E)=\frac{\Gamma_{L}\Gamma_{R}}
{(E-E_{0})^{2}+(\Gamma_{L}/2+\Gamma_{R}/2)^{2}}.
\end{equation}
Here $E_{0}$ is the energy of the localized state and
$\Gamma_{L}$, $\Gamma_{R}$ are leak rates of an electron from the
localized state to left and right leads respectively, taken as
$\Gamma_{L}\propto\exp(-2x/L_{\ell})$ and
$\Gamma_{R}\propto\exp(-2(L-x)/L_{\ell})$, where $L$ is barrier
length, $x$ is the position of impurity center and $L_{\ell}$ the
localization length. The standard frameworks for the description
of transport through a mesoscopic system are the Landauer-Buttiker
formula, methods of random-matrix theory and Kubo's formula
Ref.\cite{LK,KeldyshApproach,Beenakker}. Here we apply transfer
matrix techniques, at first suggested in Ref.\cite{LK}, with some
generalization for inclusion of inelastic processes.

When the energy of incident particle is far from the spectrum of
the random Hamiltonian, which describes the barrier with
impurities, than we are not in the resonant conditions for
tunneling. The Breit-Wigner formula is not applicable any more and
the transition probability is exponentially small
$\mathcal{T}(E)\propto\exp(-L/L^{*})$ with
$L^{*}=\hbar/\sqrt{2m(U-E)}$ where $U$ barrier hight and $L$
barrier length. The situation changes when inelastic processes are
taken into account which is relevant for low temperatures when
electron-phonon collisions become sufficiently inelastic. The
incident particle, by emission or absorbtion of a phonon, may be
trapped into the resonant level with subsequent resonant
tunneling. The probability of the inelastic scattering is
exponentially small so that total tunneling probability is also
exponentially small but it turns out that in some energy interval,
$\triangle E=|E-E_{0}|$, the inelastic tunneling probability
exponentially large compared to the elastic one in the regimes out
of resonance. The last statement may be summarized in the
following sequence of inequalities
$\mathcal{T}^{\mathrm{res}}_{\mathrm{elastic}}(E\sim E_{0})\sim
1\gg\mathcal{T}(E)_{\mathrm{inelastic}}\gtrsim\exp(-L/L^{*})$. The
aim of this work is to find out the energy interval $\triangle E$
when $\mathcal{T}(E)_{\mathrm{inelastic}}\gtrsim\exp(-L/L^{*})$
and investigate the energy dependence of tunneling probability in
that interval for various configurations of impurity centers.

\section{Formalism}
In this section we consider the tunneling of the quantum particle
with the energy $E$ through the rectangular barrier with ingrained
in it randomly distributed impurities. In the case of short-range
point-like impurities, randomly arranged in the points $z_{j}$
$(j=1,2,\ldots, N)$, the impurity potential has the form
\begin{equation}\label{Potential}
U(x)=U+K\sum_{j=1}^{N}\delta(x-z_{j}), \quad 0<z_{j}<L
\end{equation}
so that electrons are described by the single-particle Hamiltonian
\begin{equation}
\hat{H}_{0}=-\frac{\hbar^{2}}{2m}\frac{\mathrm{d}^{2}}{\mathrm{d}x^{2}}+
U+K\sum_{j=1}^{N}\delta(x-z_{j})
\end{equation}
and interact with the phonon bath maintained in the thermal
equilibrium through Fr\"{o}lich-like electron-phonon Hamiltonian.
By introducing new variables $\kappa^{2}=2m(U-E)/\hbar^{2}$,
$\mathcal{K}=2mK/\hbar^{2}$ and $k^{2}=2mE/\hbar^{2}$ we arrive at
the Schr\"{o}dinger equation
\begin{equation}\label{SchrodingerEq}
\frac{\mathrm{d}^{2}\Psi}{\mathrm{d}x^{2}}-\kappa^{2}\Psi=
\mathcal{K}\sum_{j=1}^{N}\delta(x-z_{j})
\end{equation}
with subsequent solutions of superposition the incident and
reflected wave at $x<0$ and transmitted wave at $x>L$ in the form
\begin{subequations}
\begin{equation}\label{IncidentReflectTransmitA}
\Psi(x)=\exp(\mathrm{i}kx)+r\exp(-\mathrm{i}kx), \quad x<0,
\end{equation}
\begin{equation}\label{IncidentReflectTransmitB}
\Psi(x)=t\exp(\mathrm{i}k'(x-L)), \quad x>L.
\end{equation}
\end{subequations}
where $r$ -- reflection and $t$ -- transmission coefficients.
Assuming the continuity of wave function and it derivative at the
boundaries of the barrier we have following boundary conditions
\begin{equation}\label{BoundariCond}
\Psi(0)+\frac{\Psi'(0)}{\mathrm{i}k}=2,\quad
\frac{\Psi'(L)}{\Psi(L)}=\mathrm{i}k'.
\end{equation}
In the above defined variables the the barrier transmittancy
$\mathcal{T}$ can be presented as
\begin{equation}
\mathcal{T}(E,\Gamma_{N})=|t|^{2}=|\Psi(L)|^{2}.
\end{equation}
and it depends on the energy $E$ of incident particle, number of
the impurity centers $N$ and the configuration of impurities
through the phase volume $\Gamma_{N}=\{z_{j}\}$. At this point it
is convenient to introduce new notation for description the
particle dynamics. Instead of the single wave function $\Psi(x)$
we introduce the vector-wave function
\begin{equation}
\hat{\Psi}(x)=\left(\begin{array}{c}\psi_{-}(x) \\ \psi_{+}(x)
\end{array}\right)
\end{equation}
such that
\begin{equation}
\Psi(x)=\psi_{-}(x)+\psi_{+}(x),\
\Psi'(x)=\kappa(\psi_{-}(x)-\psi_{+}(x)).
\end{equation}
Above definitions imply that the functions $\psi_{-}(x)$ and
$\psi_{+}(x)$ satisfy the following equation
\begin{equation}
\frac{\mathrm{d}\psi_{\pm}}{\mathrm{d}x}=\pm\kappa\psi_{\pm}
\end{equation}
everywhere, except the points $z_{j}$ so that
\begin{equation}
\frac{\mathrm{d}\hat{\Psi}(x)}{\mathrm{d}x}+\hat{A}\hat{\Psi}(x)=
\frac{\mathcal{K}}{2\kappa}\hat{B}\sum_{j=1}^{N}\delta(x-z_{j})\hat{\Psi}(z_{j})
\end{equation}
where
\begin{equation}
\hat{A}=\left(\begin{array}{cc}\kappa & 0 \\ 0 &
-\kappa\end{array}\right), \quad \hat{B}=\left(\begin{array}{cc}1
& 1 \\ -1 & -1\end{array}\right)
\end{equation}
and subsequent new boundary conditions
\begin{subequations}
\begin{equation}
\psi_{+}(0)(k/\kappa-\mathrm{i})+\psi_{-}(0)
(k/\kappa+\mathrm{i})=2k/\kappa,
\end{equation}
\begin{equation}
\psi_{+}(L)(k'/\kappa+\mathrm{i})+\psi_{-}(L)
(k'/\kappa-\mathrm{i})=0.
\end{equation}
\end{subequations}
Starting from now we can construct the scattering matrix $\hat{S}$
that "carries" solution through the region $0<x<L$ from the right
to the left boundary of the barrier:
\begin{equation}
\hat{\Psi}(0)=\hat{S}\hat{\Psi}(L),
\end{equation}
\begin{equation}
\hat{S}=\hat{G}(z_{1})\hat{T}_{1,2}(z^{-}_{1}|z^{+}_{1})\hat{G}(z_{2}-z_{1})\ldots
\end{equation}
\[
\ldots\hat{T}_{N-1,N}(z^{-}_{N}|z^{+}_{N})\hat{G}(L-z_{N})
\]
where $\hat{G}(x)$ is the free propagator that carries solution
from right to left into the region free from the scattering
centers
\begin{equation}
\hat{G}(x)=\left(\begin{array}{cc}\exp(\kappa x) & 0 \\ 0 &
\exp(-\kappa x)\end{array}\right)
\end{equation}
and $\hat{T}$ is the matrix that propagates solution through the
scattering center
\begin{subequations}
\begin{equation}
\hat{T}_{j,j+1}(z^{-}_{j}|z^{+}_{j})=\frac{1}{2}\left(\begin{array}{cc}
1+\frac{\kappa_{j+1}+\mathcal{K}}{\kappa_{j}} &
1-\frac{\kappa_{j+1}-\mathcal{K}}{\kappa_{j}}\\{}&{}\\
1-\frac{\kappa_{j+1}+\mathcal{K}}{\kappa_{j}} &
1+\frac{\kappa_{j+1}-\mathcal{K}}{\kappa_{j}}
\end{array}\right)
\end{equation}
\begin{equation}
\hat{\Psi}(z_{j}-0)=\hat{T}_{j,j+1}(z^{-}_{j}|z^{+}_{j})\hat{\Psi}(z_{j}+0).
\end{equation}
\end{subequations}
Also we introduce following two vectors
\begin{equation}
f=\frac{1}{2}\left(\begin{array}{c} 1+\mathrm{i}\kappa/\mathcal{K}
\\1-\mathrm{i}\kappa/\mathcal{K}\end{array}\right), \quad
g=\frac{1}{2}\left(\begin{array}{c}
1+\mathrm{i}\mathcal{K}/\kappa'
\\1-\mathrm{i}\mathcal{K}/\kappa'\end{array}\right).
\end{equation}
that make it possible to present the transition probability in the
very compact form
\begin{equation}\label{TunnelProbability}
\mathcal{T}_{N}(E,\Gamma_{N})=\int\frac{W(E,E')}{|(f,\hat{S}g)|^{2}}\mathrm{d}E'
\end{equation}
where function $W(E,E')$ represents the probability of the
electron transition from the quantum state with the energy $E$
into quantum state with the energy $E'$ as the result of
electron-phonon interaction, which may be found from linearized
kinetic equation
\begin{subequations}
\begin{equation}
W(E,E')=W^{+}(E,E')+W^{-}(E,E'),
\end{equation}
\begin{equation}\label{InelasticProbabilityAplitude}
W^{\pm}(E,E')=\int|M_{q}|^{2}\frac{1-n_{F}(E')}{1-n_{F}(E)}\frac{\delta(k(E)-\kappa(E')\mp
q)}{v(E')}
\end{equation}
\[
\times[(N_{q}+1)\delta(E-E'-\hbar\omega)+N_{q}\delta(E-E'+\hbar\omega)]\frac{\mathrm{d}q}{2\pi}
\]
\end{subequations}
where $"\pm"$ correspond to emission and absorption of the phonon,
$M_{q}$ -- matrix element of the electron-phonon interaction,
$N_{q}$ and $n_{F}$ -- boson and fermion distribution functions
and $\delta$-functions presents conservation laws of the momentum
and the energy. Representation Eq.~(\ref{TunnelProbability}) is
very useful because all information about particle dynamics in the
barrier is included in $\hat{S}$ matrix, at the same time all
information about boundary conditions lies in the vectors $f$ and
$g$, and inelastic processes included in the function $W(E,E')$.

Caring out integration in
Eq.~(\ref{InelasticProbabilityAplitude}), taking into account only
valuable exponential factors and comparing $W$ to the elastic
tunneling probability in the out of the resonance conditions
$\propto\exp(-L/L^{*})$ we can find the energy interval $\triangle
E$, where inequality
$\mathcal{T}(E)_{\mathrm{inelastic}}\gtrsim\exp(-L/L^{*})$ is
satisfied, which equals to
\begin{equation}
\triangle E\sim
T\ln\left(\frac{1}{\exp(-L/L^{*})-\exp(-|E_{0}|/T)}\right)
\end{equation}
where $T$ is temperature of the system and also assumed
$L>L^{*}(|E_{0}|/T)$.

Analyzing properties of $\hat{S}$ matrix it is possible to show
that maximum value for the tunneling probability occurs for some
particular impurity configurations (resonant configurations). The
conditions needed for $\mathrm{max}\{\mathcal{T}\}$
\begin{equation}\label{ResTunnelConditions}
X(E,\Gamma_{N})=S_{11}-S_{22}\sim 1, \quad
Y(E,\Gamma_{N})=S_{12}+S_{21}\sim 1
\end{equation}
define this resonant configuration with the statistical weight
\begin{equation}
\Gamma^{\mathrm{res}}_{N}=\int_{|X(E,z_{1}\ldots z_{N})|\lesssim
1\atop |Y(E,z_{1}\ldots z_{N})|\lesssim 1}
\mathrm{d}z_{1}\mathrm{d}z_{2}\ldots\mathrm{d}z_{N}.
\end{equation}
It ratio to the full statistical weight of all possible
configurations
\begin{equation}
\Gamma_{N}=\int_{0<z_{1}<\ldots<z_{N}<L}
\mathrm{d}z_{1}\mathrm{d}z_{2}\ldots\mathrm{d}z_{N}=\frac{L^{N}}{N!}
\end{equation}
gives the probability of the resonant configuration realization
$P_{N}(E)$ for the given energy of the incident particle
\begin{equation}
P_{N}(E)=\frac{\Gamma^{\mathrm{res}}_{N}}{\Gamma_{N}}.
\end{equation}
Averaged over all possible configuration, with the fixed number of
impurities, the resonant transmittancy has the form
\begin{equation}
\mathcal{T}_{N}(E)\approx\frac{1}{\Gamma_{N}}\int_{\Gamma^{\mathrm{res}}_{N}(E)}
\mathcal{T}_{N}(E,\Gamma_{N})\mathrm{d}\Gamma_{N}
\end{equation}
and average value of transmittancy $\langle\mathcal{T}(E)\rangle$
for arbitrary number of impurities can be found by averaging
$\mathcal{T}_{N}(E)$ with Poisson distribution function (under
assumption of absence of correlation  between impurities)
\begin{equation}\label{FullAveragedTunProbability}
\langle\mathcal{T}(E)\rangle=\sum^{\infty}_{N=1}p_{N}\mathcal{T}_{N}(E),
\quad p_{N}=(cL)^{N}\exp(-cL)/N!,
\end{equation}
where $c$ is linear impurity concentration.
\newline
\section{Tunnel transparency $\mathcal{T}(E)$ of the disordered barrier}
\textbf{A) One impurity scattering center $N=1$:} In what follows
we have introduced new parameters
$\epsilon=1+\mathcal{K}/2\kappa$, $\mathcal{L}=L\kappa$ and
$\tau=\sqrt{2mL^{2}/(U-E)}$. The $\hat{T}$ and $\hat{S}$ matrices
in the main approximation equals to
\begin{equation}
\hat{T}(z^{-}_{1}|z^{+}_{1})=\left(\begin{array}{cc}
\epsilon+\omega\tau/2 & -1-\omega\tau/2\\ 1-\omega\tau/2 &
2+\omega\tau/2
\end{array}\right),
\end{equation}
\begin{widetext}
\begin{equation}
\hat{S}=\left(\begin{array}{cc} \epsilon\exp(
\mathcal{L})\exp(\omega\tau(\mathcal{L}-\mathcal{Z}_{1})) &
-\exp(-(\mathcal{L}-2\mathcal{Z}_{1}))\exp(-\omega\tau(\mathcal{L}-\mathcal{Z}_{1}))\\
\exp(\mathcal{L}-2\mathcal{Z}_{1})\exp(\omega\tau(\mathcal{L}-\mathcal{Z}_{1}))
& 2\exp(-
\mathcal{L})\exp(-\omega\tau(\mathcal{L}-\mathcal{Z}_{1}))
\end{array}\right).
\end{equation}
The transmission probability has the form
\begin{equation}
\mathcal{T}_{1}(E,\Gamma_{N=1})\approx\frac{1}{1+
\sinh^{2}[(\mathcal{L}-2\mathcal{Z}_{1})+\omega\tau(\mathcal{L}-\mathcal{Z}_{1})]+
\frac{1}{4}\epsilon^{2}\exp(2\mathcal{L})\exp(2\omega\tau(\mathcal{L}-\mathcal{Z}_{1}))}
\frac{e^{E/T}+1}{e^{E/T}+e^{\hbar\omega/T}}\end{equation} and
subsequently
\begin{equation}
\langle\mathcal{T}_{1}(E)\rangle\approx\frac{1}{\mathcal{L}}\int_{0}^{\mathcal{L}}
\mathcal{T}_{1}(E,\Gamma_{N=1})\mathrm{d}\mathcal{Z}_{1}\approx
\frac{2} {\mathcal{L}\epsilon^{2}}
\ln\left[1+\epsilon^{2}\exp\left(4\mathcal{L}\frac{1+\omega\tau}{2+\omega\tau}\right)\right]
\exp\left(-4\mathcal{L}\frac{1+\omega\tau}{2+\omega\tau}\right)
\frac{e^{E/T}+1}{e^{E/T}+e^{\hbar\omega/T}}.
\end{equation}
\end{widetext}
\textbf{B) Two impurity scattering centers $N=2$:} considering the
case of two centers located in points $\mathcal{Z}_{1}$,
$\mathcal{Z}_{2}>\mathcal{Z}_{1}$ it is convenient to introduce
following quantities $\eta=\mathcal{Z}_{2}-\mathcal{Z}_{1}$ --
relative distance between centers and
$\zeta=(\mathcal{L}-\mathcal{Z}_{2})-\mathcal{Z}_{1}$ -- asymmetry
in the scattering centers configuration. Calculating elements of
the scattering matrix $\hat{S}$ and then the $X$ and $Y$ functions
form Eq.~(\ref{ResTunnelConditions}) we have
\begin{subequations}
\begin{equation}
X(E,\Gamma_{N=2})=[\epsilon^{2}-\exp(-2\eta(1+\omega\tau))]
\exp(\mathcal{L}-\mathcal{Z}_{1}\omega\tau)-
\end{equation}
\[
-[4-\exp(2\eta(1
+\omega\tau))]\exp(-\mathcal(L))\exp((-\mathcal{L}-\mathcal{Z}_{1})\omega\tau),
\]
\begin{equation}
Y(E,\Gamma_{N=2})=2\sinh[\zeta+(\mathcal{L}-\mathcal{Z}_{2})\omega\tau]
[\epsilon\exp(\eta(1+\omega\tau))+,
\end{equation}
\[
+2\exp(-\eta(1+\omega\tau))]
\]
\end{subequations}
\begin{figure}[tp]
\includegraphics[10,0][246,223]{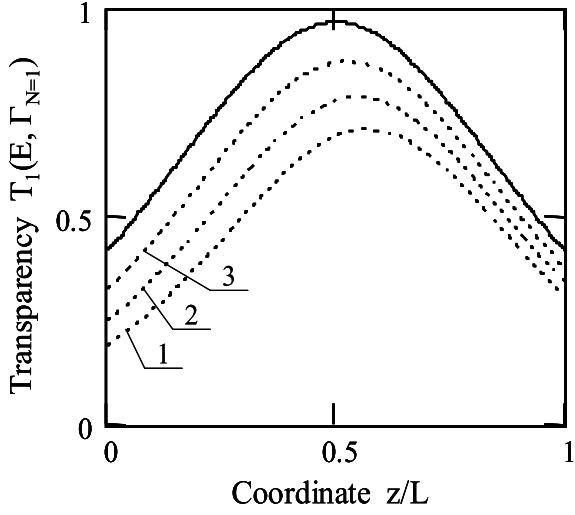}
\caption{In the graph is shown schematically the shape of the
transparency profile as the function of single impurity position
in the barrier for resonant elastic (solid line) and inelastic
(dashed lines) scattering, with differen values of energy losses
($\omega_{1}>\omega_{2}>\omega_{3}$).\label{Fig:1}}
\end{figure}
where $\Gamma_{N=2}=\{\eta,\zeta\}\Rightarrow$
$0<\eta<\mathcal{L}$ and $|\zeta|<\mathcal{L}-\eta$. Analyzing the
energy and configuration dependance of $X(E,\Gamma_{N=2})$ and
$Y(E,\Gamma_{N=2})$ it is possible to determine several
constraints, that define conditions for the maximum tunnel
probability and we list them below: a)
$\epsilon\sim\exp(-\eta(1+\omega\tau))$ and
$\epsilon\sinh[\zeta+(\mathcal{L}-\mathcal{Z}_{2})\omega\tau]\sim
1$, b) for the energy interval
$\exp(-\mathcal{L})<\epsilon<\exp(\mathcal{-L}/2)$ we have
$2\eta(1+\omega\tau)\sim\mathcal{L}$ and
$\sinh[\zeta+(\mathcal{L}-\mathcal{Z}_{2})\omega\tau]\sim\epsilon^{-1}\exp(-\mathcal{L}/2)$,
c) $2\eta(1+\omega\tau)\sim\mathcal{L}$ and
$\zeta+(\mathcal{L}-\mathcal{Z}_{2})\omega\tau\sim 1$ for the
energy $\epsilon\sim\exp(-\mathcal{L}/2)$ and d)
$\epsilon\gg\exp(-\mathcal{L}/2)$ independently on configuration.
By knowing $\eta$ and $\zeta$ we can determine the probability for
two impurity resonant configuration formation
\begin{equation}
P_{N=2}(E)=\frac{2}{\mathcal{L}^{2}}\int_{
0<\eta<\mathcal{L}\atop|\zeta|<\mathcal{L}-\eta}
\mathrm{d}\eta\mathrm{d}\zeta
\end{equation}
and subsequently derive the resonant barrier transmittancy
\begin{equation}
\mathcal{T}_{2}(E)\approx\left\{\begin{array}{cc}
\left|\ln\epsilon+\mathcal{L}\frac{\omega\tau}{2+\omega\tau}\right|\frac{2\mathcal{F}(E)}{\mathcal{L}
[1+(2-\epsilon^{2}/2)\exp(-\mathcal{L})]}&
(\mathrm{a})\\{}\\
\frac{\ln[\epsilon\exp{\mathcal{L}/2}]}{2+\omega\tau}
\frac{2\mathcal{F}(E)}{\mathcal{L}(1+\exp(-\mathcal{L}/2))} & (\mathrm{b}) \\{}\\
\left(\frac{1-2\omega\tau}{1+2\omega\tau}\right)\frac{\mathcal{F}(E)}{1+2\exp(-\mathcal{L})}
& (\mathrm{c})\\{}\\
\frac{2\mathcal{F}(E)}{\epsilon^{2}\mathcal{L}^{2}}
\exp\left(-\mathcal{L}\frac{1+\omega\tau}{1-\omega\tau}\right) &
(\mathrm{d})
\end{array}\right.
\end{equation}
for $(\mathrm{a})-(\mathrm{d})$ cases described above
subsequently, with
$\mathcal{F}(E)=(e^{E/T}+1)/(e^{E/T}+e^{2\hbar\omega/T})$. Also it
is possible to estimate asymptotic behavior of the averaged
barrier transparency for all possible impurity configurations
\begin{equation}
\langle\mathcal{T}_{2}(E)\rangle\sim\frac{1}{\mathcal{L}^{2}}\exp\left[-\frac{\mathcal{L}}{2}
\left(\frac{1+\omega\tau}{1-\omega\tau}\right)\right]
\frac{e^{E/T}+1}{e^{E/T}+e^{2\hbar\omega/T}}.
\end{equation}
\begin{figure}[here]
\includegraphics[60,0][316,255]{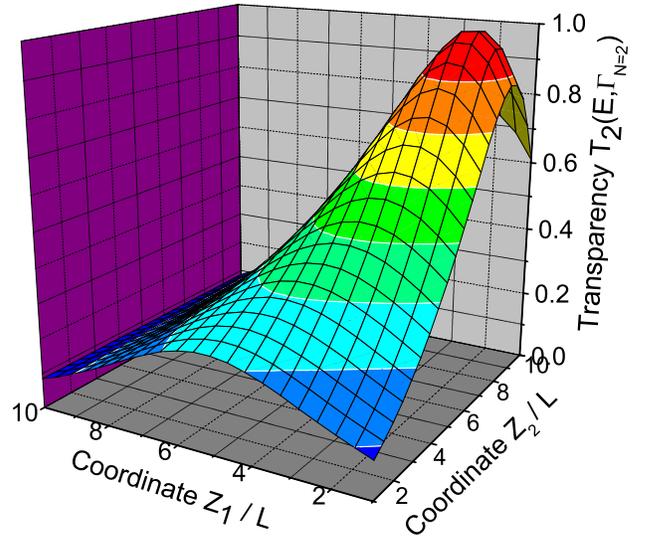}
\caption{One example of the shape for the transparency
$\mathcal{T}_{2}(E,\Gamma_{N=2})$ profile as the function of
impurity positions $\mathcal{Z}_{1}$ and $\mathcal{Z}_{2}$. The
distinctive maximum correspond to resonant configuration of type
$(\mathrm{c})$, by classification scheme for two-impurity
tunneling, when the energy of the incident particle
$\epsilon\sim\exp(-\mathcal{L}/2)$. In the $x$ and $y$ axes is
shown the ratio $\mathcal{Z}/\mathcal{L}$ in pixel such that 10
pixels correspond to ratio $\mathcal{Z}/\mathcal{L}=1$ and in the
$z$ axis is shown absolute value of the tunnel
probability.\label{Fig:2}}
\end{figure}
\textbf{C) General case of $N$ impurity centers arranged as
$0<\mathcal{Z}_{1}<\ldots<\mathcal{Z}_{N}<\mathcal{L}$:} in this
case detailed analysis on the basis of above suggested formalism
is very difficult, but rough estimation for the tunneling
probability $\langle\mathcal{T}_{N}(E)\rangle$ may be found. It
turns out that the picture of tunneling under conditions of large
$N\gg 1$ breaks into effectively two-centers problem. This is true
if among all centers there are two, which separated as $\delta
\mathcal{Z}=\mathrm{max}\{\mathcal{Z}_{j+1}-\mathcal{Z}_{j}\}\sim|\ln|\epsilon||$.
Such constraint comes from the fact that the $S_{11}=\epsilon
e^{\delta\mathcal{Z}}$ matrix element of the scattering matrix is
exponentially dominant and if in some energy interval we can put
$S_{11}\sim 1$ than tunneling occur with maximum probability,
otherwise it exponentially small. Subsequently, tunnelling via
these centers dominate the whole tunneling process. The main
contribution to the tunneling probability
$\langle\mathcal{T}_{N}(E)\rangle$ in summation formula
Eq.~(\ref{FullAveragedTunProbability}) comes form optimal $N$ such
that $N=N_{\mathrm{opt}}\sim\sqrt{\mathcal{L}/|\ln c|}\gg 1$. The
resonant impurity configuration, with mentioned constraint for
$\delta\mathcal{Z}$, is in the order of
$\Gamma^{\mathrm{res}}_{N}\sim
2^{-N}(N\delta\mathcal{Z}-\mathcal{L})^{N-1}/(N-1)!$ and
asymptotic for the tunneling probability
\begin{equation}
\langle\mathcal{T}_{N}(E)\rangle\sim\exp\left(-2\sqrt{\mathcal{L}|\ln
c|}-\mathcal{L}c-(\hbar\omega/T)\sqrt{\mathcal{L}/|\ln c|}\right).
\end{equation}

To summarize, we have considered the tunneling of the quantum
particle through disordered barrier in the regime of inelastic
scattering when the tunneling probability is greater, compared to
that of elastic scattering, in the out of resonance conditions.
The energy interval $\triangle E$ where such situation may be
realized is found and set of the tunneling probability functions
$\mathcal{T}(E)$ for various impurity configurations presented.
\newline

\end{document}